\documentclass[%
 reprint,
 amsmath,amssymb,
 aps,
]{revtex4-1}

\usepackage[utf8]{inputenc}
\usepackage[english]{babel}
\usepackage{graphicx}
\usepackage{dcolumn}
\usepackage{bm}
\usepackage{hyperref}

\begin{document}

\preprint{APS/123-QED}

\title{Nonequilibrium thermodynamics of input-driven networks}

\author{Kevin S. Chen}
  \email{kschen@princeton.edu}
\affiliation{%
Princeton Neuroscience Institute, Princeton University, Princeton, NJ, USA
}%




\date{\today}

\begin{abstract}
Neural dynamics of energy-based models are governed by energy minimization and the patterns stored in the network are retrieved when the system reaches equilibrium. However, when the system is driven by time-varying external input, the nonequilibrium process of such physical system has not been well-characterized. Here, we study attractor neural networks, specifically the Hopfield network, driven by time-varying external input and measure thermodynamic quantities along trajectories between two collective states. The overlap between distribution of the forward and reversal work along the nonequilibrium trajectories agrees with the equilibrium free energy difference between two states, following the prediction of Crooks fluctuation theorem. We study conditions with different stimulation protocol and neural network constraints. We further discuss how biologically plausible synaptic connections and information processing may play a role in this nonequilibrium framework. These results demonstrate how nonequilibrium thermodynamics can be relevant for neural computation and connect to recent systems neuroscience studies with closed-loop dynamic perturbations.

\begin{description}
\item[Keywords]
Nonequilibrium thermodynamics, Hopfield network, Fluctuation theorem, Attractor neural networks, Hebbian learning
\end{description}
\end{abstract}

\pacs{Valid PACS appear here}
\maketitle


\section{\label{sec:level1}Introduction}

Neural network dynamics are fundamental for neural computation, decision-making, and learning and memory \cite{Vogels2005-sz,Hertz2018-yw,Rajan2016-ck,Frady2019-in}. Attractor models have been used to study neural recordings in modern neuroscience experiments, including fixed-points, oscillations, and chaotic dynamics \cite{Vogels2005-sz,Hopfield1982-hu,Sompolinsky1988-yy,Seung1996-aq, Tank1987-oe}. These analyses rely on repeating trials and search for stationary neural activity patterns in absence of dynamical stimuli \cite{Hopfield1982-hu, Rajan2010-pd,Seung1993-mn}. However, real neural networks are often driven by time-varying stimulus.
Neural responses to stimuli have been classically characterized with tuning curves and recently extended to population encoding, but the connections to attractor neural dynamics are less clear \cite{Pillow2008-vg,Seung1993-mn}. Here, we study input-driven neural networks in a theoretical framework that can be analyzed by nonequilibrium thermodynamics.

The Hopfield model provides a framework to describe neural dynamics in terms of an energy function or Lyapunov function for dynamical systems \cite{Hopfield1982-hu,Hopfield1984-je}. The resulting neural dynamics asymptotically reaches a stationary state known as the attractor in dynamical systems. In terms of thermodynamics, this is the result when the system reaches thermal equilibrium and the probability of activity patterns follows Boltzmann distribution \cite{Hertz2018-yw,Hinton2012-vu}. Specifically, these equilibrium solutions are learned through synaptic learning rules and the recalling process of learned patterns captures the property of associative memory \cite{Hertz2018-yw}. In more biological settings, the neural network receives time-varying external input and the synaptic weights may change on the time-scale that is not separable from neural dynamics. This leads to nonequilibrium conditions as the system is constantly driven away from equilibrium neural states. Stability analysis and Boltzmann distribution for equilibrium systems are not relevant for analyzing dynamics far-from-equilibrium. 

Nonequilibrium thermodynamics studies state variables that are away from equilibrium and how macroscopic variables used in equilibrium thermodynamics can be described under such conditions \cite{Jarzynski1997-mg,Jarzynski2011-lw}. In the past few decades, the relation between nonequilibrium work distribution and equilibrium free energy difference has been theorized and supported experimentally \cite{Crooks1998-ie, Jarzynski1997-mg, Collin2005-fc, Cetiner2020-ci, Seif2020-ht}. More specifically, Crooks theory relates the ratio between forward and backward processes with dissipated work, as well as providing a nonequilibrium method for measuring free energy difference between states \cite{Crooks1998-ie,Crooks1999-zg}. Fluctuation theory enables thermodynamic characterization of systems that are stochastic and driven by external input or time-varying parameters, such as biological neural networks. Characterizing neural activity in terms of free energy is of interest as it informs kinetics, relates to thermodynamics properties, and further tests for theories of brain computations \cite{Qian2005-tm, Hertz2018-yw, Friston2010-es}.

Recently, nonequilibrium statistical mechanics has been applied to input-driven continuous attractors \cite{Zhong2018-vz, Yan2013-os} and unsupervised learning of restricted Boltzmann machine \cite{Salazar2017-xr}. The former derives limits of attractor dynamics and memory capacity for a bump attractor model under dynamical input, whereas the later characterizes the learning process of energy models with thermodynamic variables. For supervised learning process, the thermodynamic efficiency of different learning rules have been computed in a stochastic thermodynamics framework \cite{Goldt2017-iy}. This thread of theoretical work provides better understanding for the nonequilibrium nature of neural dynamics and learning processes. In this manuscript, we extend the picture by applying generalized fluctuation theorem to energy based neural dynamics driven by external input.

We focus on stochastic Hopfield network, also known as a Boltzmann machine \cite{Hertz2018-yw, Hinton2012-vu}, driven by external stimuli switching between two stored activity patterns. We measure work performed during the stimulus protocol and determine the equilibrium free energy according to work fluctuations. We show that the system satisfies Crooks fluctuation theorem (CFT) under dynamic input and heat exchange fluctuation theorem (XFT) in contact with heat reservoir with another temperature. Finally, we try to generalize the method to continuous systems driven by stimuli and with more biological neural networks. We discuss how nonequilibrium thermodynamics offer insight to input-driven neural dynamics and computation.

\section{\label{sec:level2}Settings}

We start by introducing nonequilibrium thermodynamical measurements through fluctuation theorems. We formulate the toy example of input-driven stochastic Hopfield network and specify thermodynamics quantities in this model. We measure work performed during a set of stimulation protocol and further extend the method with other parameter choices.

\subsection{\label{sec:level1} Fluctuation theorems}
Laws of thermodynamics can be derived from microscopic descriptions in statistical mechanics at equilibrium. Considering an isothermal process moving from an initial equilibrium state to a final equilibrium state, the second law of thermodynamic suggests that the work performed during this process $W$ and free energy difference between two states $\Delta F$ follows $W \geq \Delta F$. Free energy is defined by $F=\langle E \rangle-TS$, where $E$ is the internal energy, $T$ is temperature, and $S$ is the entropy at equilibrium. When the process between two states is nonequilibrium, Jarzynski equality states the average work performed is still related to the equilibrium free energy difference \cite{Jarzynski1997-mg}:

\begin{equation} \label{JE}
\langle\exp{(-\beta W)} \rangle = \exp{(-\beta \Delta F)},
\end{equation}

where $\beta=1/k_BT$ is the inverse temperature $T$ and $k_B$ is the Boltzmann constant. This leads to a similar form in statistical mechanics: $\langle W\rangle \geq \Delta F$, according to Jensen's inequality. For reversible process, the equality $\langle W \rangle = \Delta F$ holds, so one can calculate the free energy difference from the average of nonequilibrium work trajectories. However, for irreversible processes with dissipated work $W_{diss} = \langle W \rangle - \Delta F \neq 0$, this method of estimating free energy difference leads to error when the system is far from equilibrium. With work parameters that evolve slowly, it is shown that the work distribution can be approximated as Gaussian distribution. This leads to $\Delta F = \langle W \rangle - \beta \frac{\sigma_W^2}{2}$, where $\sigma_W^2$ is the standard deviation of work distribution \cite{Jarzynski2011-lw}. The result agrees with fluctuation-dissipation theorem with $W_{diss} = \beta \frac{\sigma_W^2}{2}$.

Crooks fluctuation theorem is a generalized form that relates forward and reverse work processes with free energy \cite{Crooks1999-zg}. Given a controllable external input parameter $I$ that controls the switching between two states in a system, the work $W$ distribution for forward $P_f$ and reverse $P_r$ is related to the equilibrium free energy difference:

\begin{equation} \label{CFT}
\frac{P_f(W)}{P_r(-W)} = \exp{(\beta(W-\Delta F))}
\end{equation}

This fluctuation theorem implies that under certain control protocol $I$, one can analyze fluctuation of nonequilibrium work trajectories and identify $W^*$ so $P_f(W^*)=P_r(W^*)$, leading to $W^*=\Delta F$. This is the nonequilibrium work method of measuring equilibrium free energy difference between two states. 

Heat exchange between two systems without any work performed follows the heat exchange fluctuation theorem \cite{Jarzynski2004-ml}. After a system reaches equilibrium at temperature $T_1$, it is placed in thermal contact with another heat reservoir at a different temperature $T_2$. Heat exchanged between two system follows a similar form:

\begin{equation} \label{XFT}
\frac{P(+ \Delta Q)}{P(-\Delta Q)}=\exp{(\Delta \beta\Delta Q)}
\end{equation}

where $+\Delta Q$ and $-\Delta Q$ are heat transferred in two directions and $\Delta \beta = 1/T_1 - 1/T_2$. With the definition of internal energy $E=W+Q$, we can replace $\Delta Q$ variables with $\Delta E$ in XFT under the condition without work performed $W=0$.

\subsection{\label{sec:level1}Input driven stochastic Hopfield network model}
Neural activities for each neuron $i$ in the Hopfield network are binary spins $V_i=\left\{+1,-1\right\}$ and neuron $j$ to neurons $i$ is connected with synaptic weight $W_{ij}$ (Fig. \ref{fig1}a) \cite{Hopfield1984-je}. The energy function can be written as a function of neural activation patterns and external input $I_i$:

\begin{equation} \label{energy}
E(V,I) = -\frac{1}{2}\Sigma_i\Sigma_j W_{ij}V_iV_j - \Sigma_j V_iI_i + \Sigma_i U_i V_i
\end{equation}

where $U_i$ is the value of neuron $i$ that sets an activation value for the net input $H_i$:

\begin{equation} \label{H}
H_i = \Sigma_j W_{ij} V_j + I_i
\end{equation}

which leads to neural dynamics through time $t$ with discrete time steps:

\begin{equation} \label{dynamics}
V_i(t+1) = \sigma(H_i(t)-U_i)
\end{equation}

where $\sigma$ is the a nonlinear function that maps to a probability of flipping $V_i$ to $+1$ or suppressing at $-1$. Here we use the sigmoid function for be consistent with the mean field approximation of spin glass models: 

\begin{equation} \label{sigmoid}
\sigma(x) = \frac{1}{1+\exp(-\beta x)}
\end{equation}

Note that this form of neural dynamics in equation \ref{H}-\ref{sigmoid} agrees with the minimization of energy function in equation \ref{energy}. Simulation of such energy-based neural dynamics is known is Glauber dynamics. At thermal equilibrium, the probability of finding neural activity given a static external stimulus follows the Boltzmann distribution:

\begin{equation} \label{Boltzmann}
P(V) = \frac{1}{Z_I} \exp(-\beta E(V,I))
\end{equation}

where the partition function $Z_I = \Sigma_V \exp(-\beta E(V,I))$. Associative memory patterns should be retrieved at thermal equilibrium. To build in memory of patterns in the Hopfield network, namely the local minimum in the energy landscape, we apply Hebbian learning rule to the connectivity matrix $W_{ij}$ \cite{Hopfield1982-hu}:

\begin{equation} \label{Hebb}
W_{ij} = \frac{1}{M} \Sigma^M_\mu \epsilon^\mu_i \epsilon^\mu_j
\end{equation}

where $M$ patterns are learned and each pattern $\epsilon^\mu$ is a binary vector with the length of $N$ neurons in the network. Note that this learning rule produces a symmetric matrix $W_{ij}=W_{ji}$ and we remove self-connections $W_{ii}=0$, as done in the original Hopfield model.

Importantly, we can now define work $W$ performed by controlling the external input $I$ and heat production through relaxation of $V$ dynamics.

\begin{equation} \label{work}
W = \Sigma^{\tau-1}_{t=0} E(V_t,I_{t+1}) - E(V_t,I_t)
\end{equation}
\begin{equation} \label{heat}
Q = \Sigma^\tau_{t=1} E(V_t,I_t) - E(V_{t-1},I_t)
\end{equation}
\begin{equation} \label{dE}
\Delta E = W + Q
\end{equation}

where the time evolution from $0$ to $\tau$ follows the time-varying stimuli protocol $I(t)$. Note that we integrate work performed through the process for multiple instantiations to compute the distribution of work $W$. With equation \ref{energy} and \ref{work}, we derive that $W = \Sigma^\tau_t \delta w_t$, where $\delta w_t = \delta I V(t)$ with $\delta I$ being the change in dynamic input signal. Note that this is consistent with electrical circuits, where power is defined by the injected current multiplying voltage $P=IV$.

\subsection{\label{sec:level1}Model settings and stimulus protocol}

\begin{figure}[t!]
    \centering
    \includegraphics[width=0.45\textwidth]{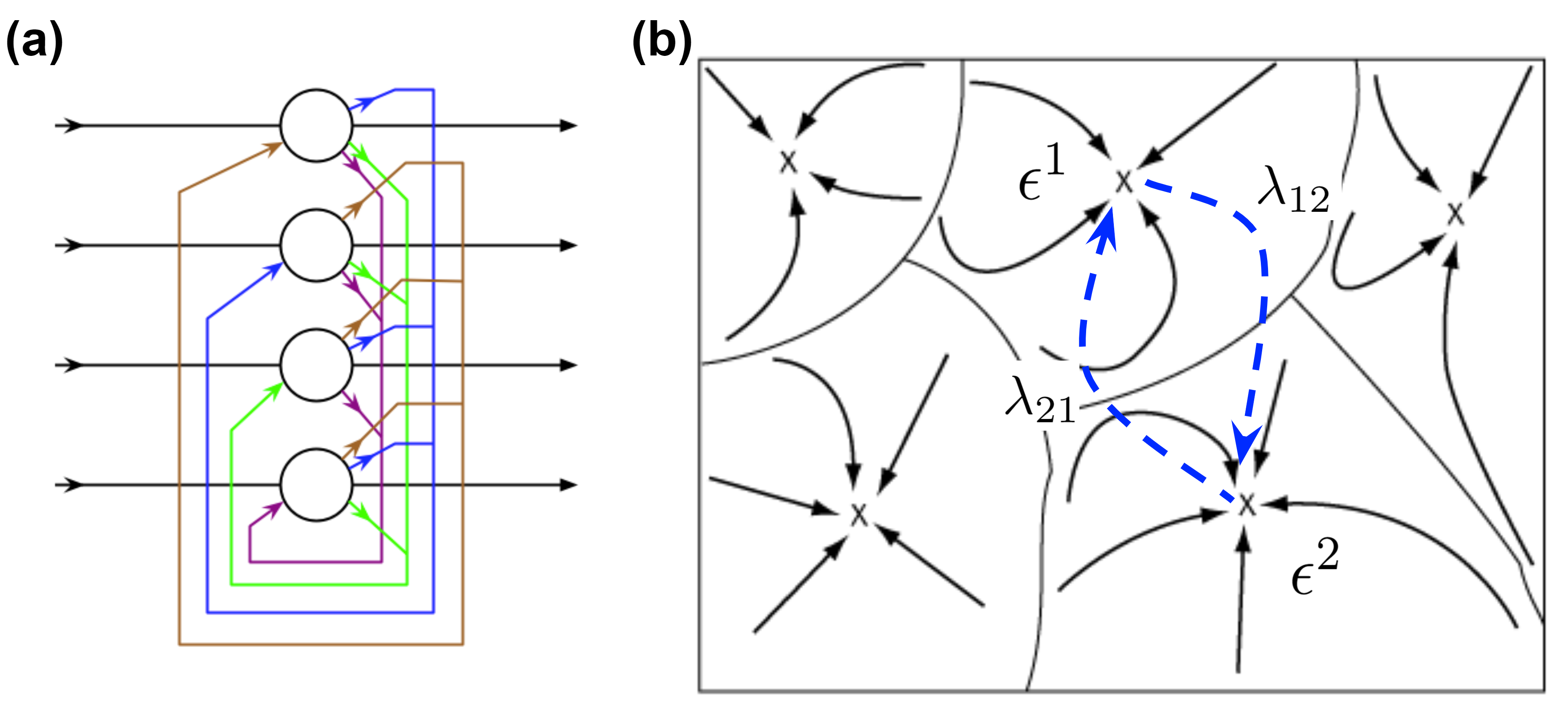}
    \caption{Model setting and input-driven network dynamics. (a) A schematic for Hopfield network. White circles are neurons and the arrows show direction of synaptic connections. Input is applied from the left, entering the recurrent network, then forming output on the right. (b) Flow trajectories in the neural state-space (black arrows) point towards attractor states (crosses). The work protocol drives the system between two equilibrium states $\epsilon_1$ and $\epsilon_2$. The forward and backward protocols $\lambda_{12}$ and $\lambda_{21}$ are shown in blue dashed arrows. Schematic modified from: \url{https://en.wikipedia.org/wiki/Hopfield_network}.}
    \label{fig1}
\end{figure}

We construct a simple Hopfield network with two memory patterns. The factor $\gamma$ that controls the weighting of memory is introduced to provide asymmetry between to stable states $\epsilon^1$ and $\epsilon^2$:

\begin{equation} \label{ww}
W_{ij} = \gamma \epsilon^1_i \epsilon^1_j + (1-\gamma)\epsilon^2_i \epsilon^2_j +\alpha \epsilon^1_i \epsilon^2_j
\end{equation}

where the $0 \leq \gamma \leq 1$ and $0 \leq \alpha \leq 1$. Note that $\gamma$ affects the energy function, thereby changing the probability of recovering two patterns at equilibrium according to equation \ref{Boltzmann}. The parameter $\alpha$ weights the overlapping between two patterns and when this it is non-zero the weight matrix can be asymmetric.  This is the original Hopfield model with two learned patterns with $\gamma=0.5$ and when $\alpha=0$. We design a stimulation protocol that drives the system away from equilibrium. This time-varying input switches between two states (Fig. \ref{fig1}b). The initial state is defined as $\epsilon^1$ and the other as $\epsilon^2$. The forward process is $\lambda_{12}: \epsilon^1 \rightarrow \epsilon^2$ and reverse is $\lambda_{21}:\epsilon^2 \rightarrow \epsilon^1$. The input protocol follows:

\begin{equation} \label{I(t)}
    I(t)=
    \begin{cases}
      \frac{st}{t_s}\epsilon^2, & \text{forward} \\
      \frac{st}{t_s}\epsilon^1, & \text{reverse}
    \end{cases}
\end{equation}

where $s$ is a factor that controls the rate of input strength increasing through time $t$ and $t_s$ is the time course of stimuli to normalize the time length of process. We compute the work performed along two processes with equation \ref{work} and compare with the equilibrium free energy difference by $\Delta F = \Delta  \langle E \rangle - T \Delta S$, where the internal energy is $\langle E \rangle=\Sigma_V P(V)E(V)$ and entropy is $S = -\Sigma_VP(V)\log P(V)$ with $P(V)$ given by equation \ref{Boltzmann}. For ground truth of $\Delta F$ between states, we numerically compute the value through $F=-\beta^{-1}\log Z$, where $Z$ is the partition function in equation \ref{Boltzmann}.

For the following numerical analyses, we use the same set of parameters, if not further noted. Network parameters: $N=15$, $k_BT=15$, $U=0$, $\alpha=0$, $\gamma =0.2$; Input protocol: $s=0.005$ with maximum iteration steps equals $1000$. Stored patterns $\epsilon^1$ and $\epsilon^2$ are randomly generated for each instantiation and the protocols are repeated for more than $5000$ repetitions for statistical measurements.

\subsection{\label{sec:level1}Biophysical neural network models}

We can extend simple binary spiking process to graded response with a continuous input-output relation that mimics biophysical properties observed in experiments \cite{Hopfield1984-je}. For continuous dynamics, the energy function can be modified as:

\begin{equation} \label{continue}
E(V,I) = -\frac{1}{2}\Sigma_i\Sigma_j W_{ij}V_iV_j - \Sigma_j V_iI_i + \Sigma_i \frac{1}{R_i}\int^{V_i}_0 f^{-1}_i(V)dV
\end{equation}

where $f$ is an input-output monotonic smooth function that has an inverse function $f^{-1}$ and $R_i$ is the effective resistance of the neurons $i$. This leads to continuous dynamics $\frac{du_i}{dt} = -\frac{u_i}{R_i} + \Sigma_jW_{ij}V_j + I_i + \eta_i$, where variable $u_i = f^{-1}_i(V_i)$ and $\eta_i$ is a noise term ($\langle \eta_i(t) \rangle = 0$, $\langle \eta_i(t)\eta_i(t') \rangle = 2B\delta (t-t')$, with fluctuating strength $B \propto k_B T$). This is a general form for recurrent neural networks \cite{Vogels2005-sz, Hertz2018-yw}. In terms of biophysics, $u$ can be thought as the continuous subthreshold voltage and $V$ can be thought as the discrete spikes. When the function $f$ is a step function, this recovers the dynamics of equation \ref{sigmoid} at high temperature. We assume the nonlinear function is a sigmoid curve $f(x) = \frac{1}{1+\exp{(-x)}}$ and take the output as an approximation of firing rate $r(t) = f(u(t))$. The firing rate $r(t)$ can be viewed as a low-pass filtered $V(t).$

Furthermore, biological learning processes are not strictly separated in time. Neural networks do not finish the learning protocol as equation \ref{Hebb} before neural dynamics in equation \ref{dynamics}. The weight matrix can be a time-varying variable that changes through time:

\begin{equation} \label{synaptic}
\frac{dW_{ij}}{dt} = -\frac{1}{\tau_s}W_{ij} + \eta V_i V_j
\end{equation}

where $\tau_s$ is the synaptic time constant and the second term on the right hand side is similar to the Hebbian learning rule in equation \ref{Hebb}. This term follows the associative learning rule with a learning rate $\eta$. The steady-state solution for the connectivity matrix $W_{ij}$ would be the time average of input patterns at long time-scale, agreeing with the form in equation \ref{Hebb}.

We construct the network connectivity $W_{ij}$ with the same method shown in discrete cases. However, the continuous firing rate output can vary between two target patterns. We compute the correlation between firing pattern and the target $\langle \epsilon^\mu r(t) \rangle$ for both patterns across all neurons. The time series of correlation across $N$ neurons are used to fit a hidden Markov model (HMM). There are two discrete states for this HMM and the emission is Gaussian. The states $m^{\mu}$ assigned by HMM represents the matching discrete pattern. The off diagonal terms along the transition matrix inferred from this HMM would be the transition probability between two states. We can then calculate an effective kinetic rate between two target states.

\section{\label{sec:level3}Results}

\subsection{\label{sec:level1}Work protocol and distribution}
\begin{figure}[b!]
    \centering
    \includegraphics[width=0.5\textwidth]{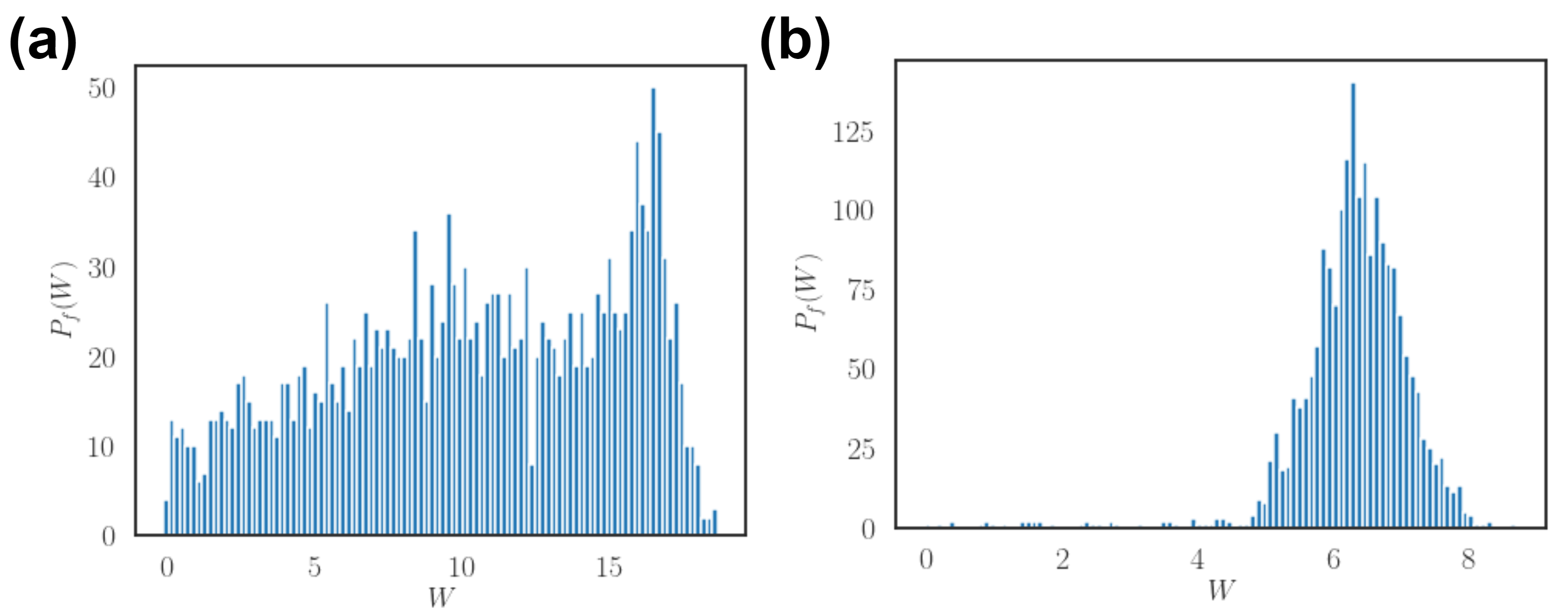}
    \caption{Work distribution in input-driven networks. Distribution of forward work value under slope factor in the input protocol (a) $s=0.005$ and (b) $s=0.05$.}
    \label{fig2}
\end{figure}

\begin{figure}[b!]
    \centering
    \includegraphics[width=0.5\textwidth]{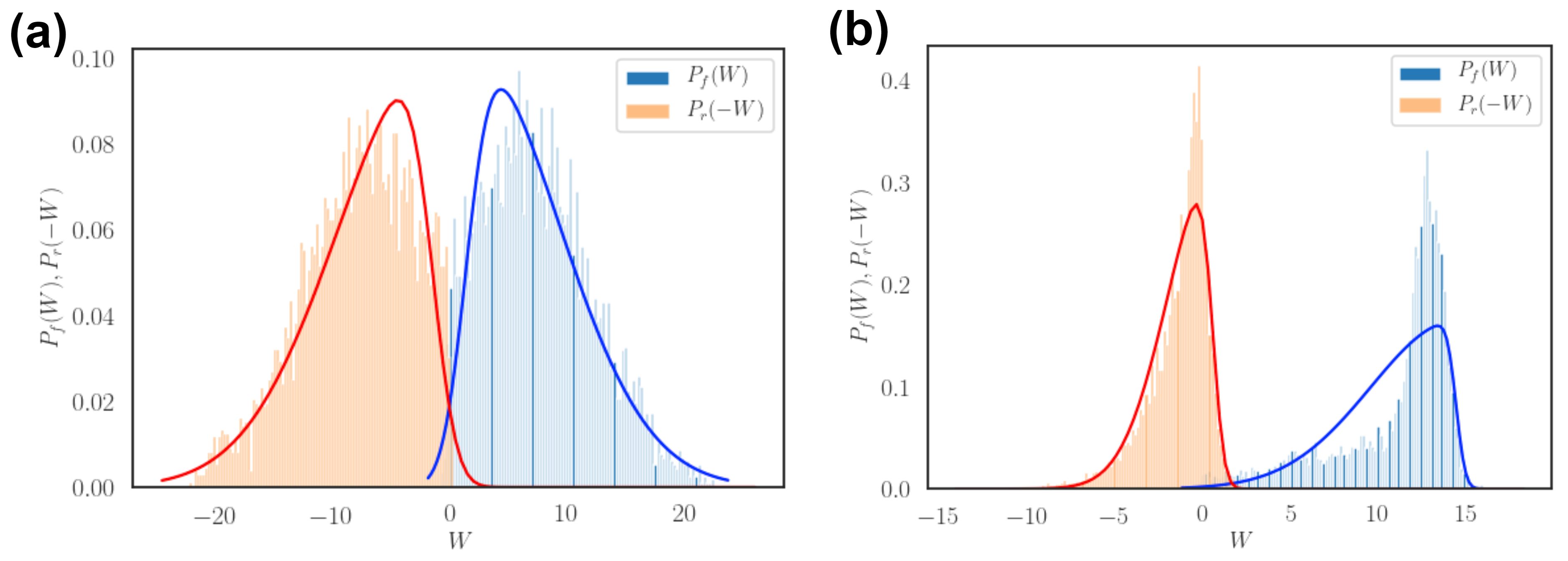}
    \caption{Forward and backward work distributions. Work distribution under networks with (a) $\Delta F=0.1$ and (b) $\Delta F=1.2$. Curves are skewed Gaussian fit to distributions.}
    \label{fig3}
\end{figure}

We measure work performed along trajectories simulated across different random instantiates with the same network parameter and nonequilibrium protocol. The distribution of work is non-Gaussian and has skewed tails or multi-modal, depending on the network and work parameters (Fig. \ref{fig2}). The multi-modal cases correspond to patterns with "glassy" energy landscape and have multiple local minimum, resulting in different modes of work required. The shape of distribution would not affect the measurement of equilibrium free energy difference according to equation \ref{CFT}. We can measure the value of $W^*$ at $P_f(W^*)=P_r(W^*)$ by searching for the crossing point of two distributions (Fig. \ref{fig3}). The value roughly agrees with true equilibrium $\Delta F$ calculated numerically. In addition to directly plotting the intersection of two distributions (Fig. \ref{fig4}a,c), we apply sampling method based on Bennett's acceptance ratio to solve for $\Delta F$ (Fig. \ref{fig4}b,d). This method does not rely on histogram binning and minimizes sampling variance \cite{Collin2005-fc, Cetiner2020-ci}. We show that the result agrees with histogram overlapping method and the ground truth free energy value. We systematically compare these values in the next section.

The result is not sensitive to the choice of work parameter within a range nor the memory patterns stored in the network. Note that the rate of forward and reverse work protocols would change the width of work distributions. When the changing rate is sufficiently slow, it is near equilibrium and the distribution concentrates near the equilibrium free energy. Similarly, the temperature of the system also governs the work distribution. Higher temperature corresponds to noisier trajectories, resulting in broader distributions or longer tails.

\subsection{\label{sec:level1}Verifying fluctuation theorems}
\begin{figure}[b!]
    \centering
    \includegraphics[width=0.5\textwidth]{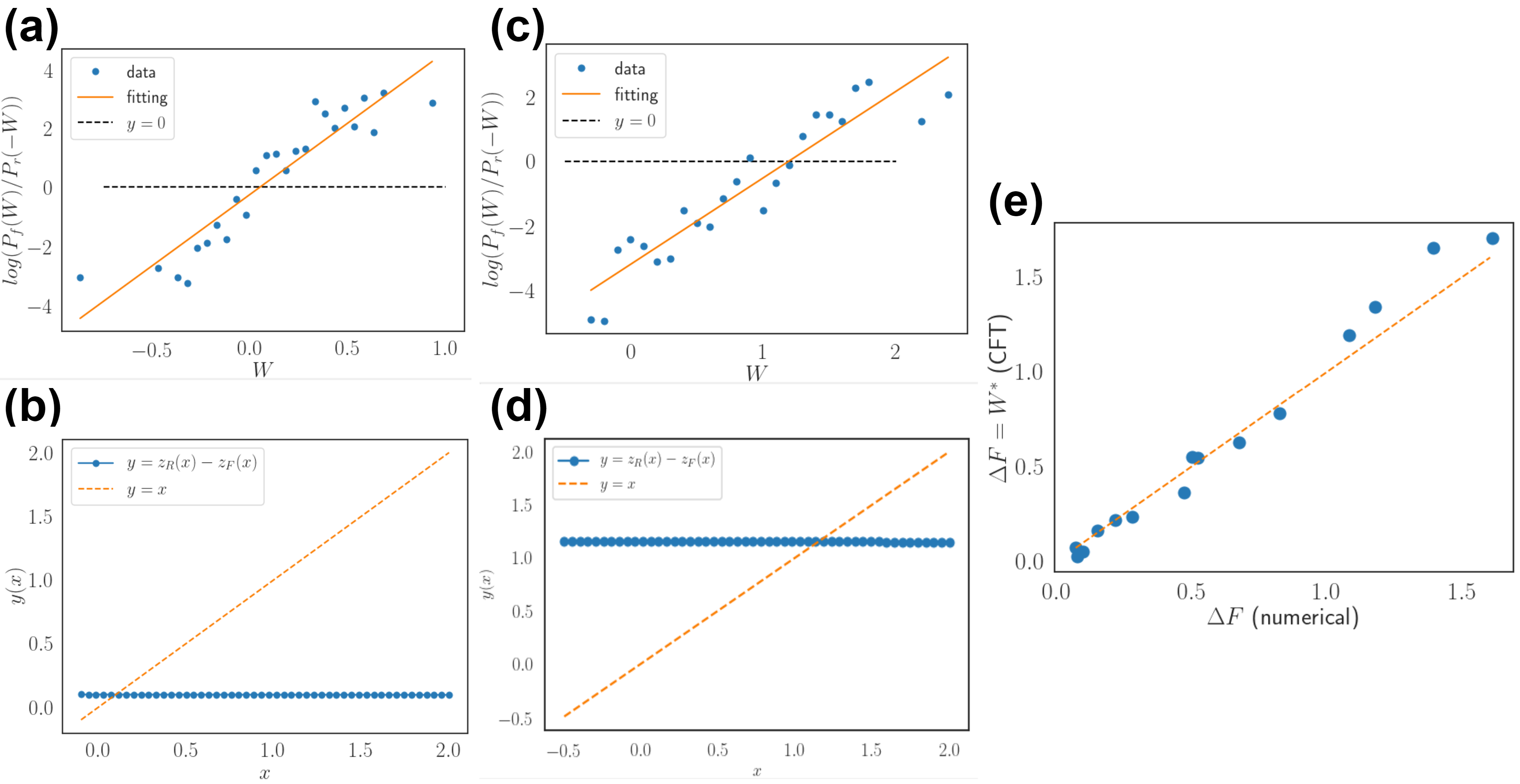}
    \caption{Verifying CFT. (a,c) Measuring the crossing point of two distributions $P_f(W)$ and $P_r(-W)$ as a function of work value $W$. Another method is through BAR (b,d), where we sample with a function $f_x(W)=(1+\exp{(\beta (x-W))}^{-1}$. The optimal $W^*=\Delta F$ value is then the crossing of function $y(x)=z_R(x)-z_F(x)$ and $x=x$, where $z_F=\log(\langle f_x(W)\exp{(-\beta W)} \rangle_F$ and $z_R=\log(\langle f_x(W) \rangle_R$. The true network $\Delta F=0.49$ for (a,b) and $\Delta F=1.25$ for (c,d). A systematic evaluation by constructing networks with different $\Delta F$ through tuning $\gamma =0.1-0.9 $ is shown in (e). Dash-line is when the numerical results agrees with our measurement through CFT.}
    \label{fig4}
\end{figure}

We systematically vary the parameter $\gamma$ in equation \ref{ww} to change the equilibrium free energy difference $\Delta F$ and estimate it through nonequilibrium work protocols shown in the former section. The results show that the nonequilibrium method recovers the value of $\Delta F$ within some range, supporting the original prediction from CFT (Fig. \ref{fig4}e). 

This "range" of parameters suitable for CFT has to suffice certain conditions: (i) tails of two distribution should overlap and provide unambiguous crossing point, (ii) fluctuation of the finite measurements has to be smaller than the scale of true equilibrium free energy difference, and (iii) the work protocol should not act too fast and dominating neural dynamics. Following these criteria, this method fails when the temperature is too high, the network is "glassy" and stored with multiple patterns, or when the work distribution is not sufficiently sampled.

Note that the well-known result following CFT is Jarzynski equality, but this method of free energy recovery fails in our setting. As shown in equation \ref{JE}, one can compute the free energy difference from nonequilibrium work trajectories. However, as noted in previous literature \cite{Collin2005-fc}, Jarzynski equality is less applicable to more dissipative processes such as the driven neural network model. Taking the logarithm of mean exponential values also leads to statistical errors. The result of Jarzynski equality is recovered when we apply external input at a much slower rate. Agreeing with previous results \cite{Collin2005-fc}, CFT provides a general method for free energy estimation with arbitrary work protocols.

\begin{figure}[t!]
    \centering
    \includegraphics[width=0.5\textwidth]{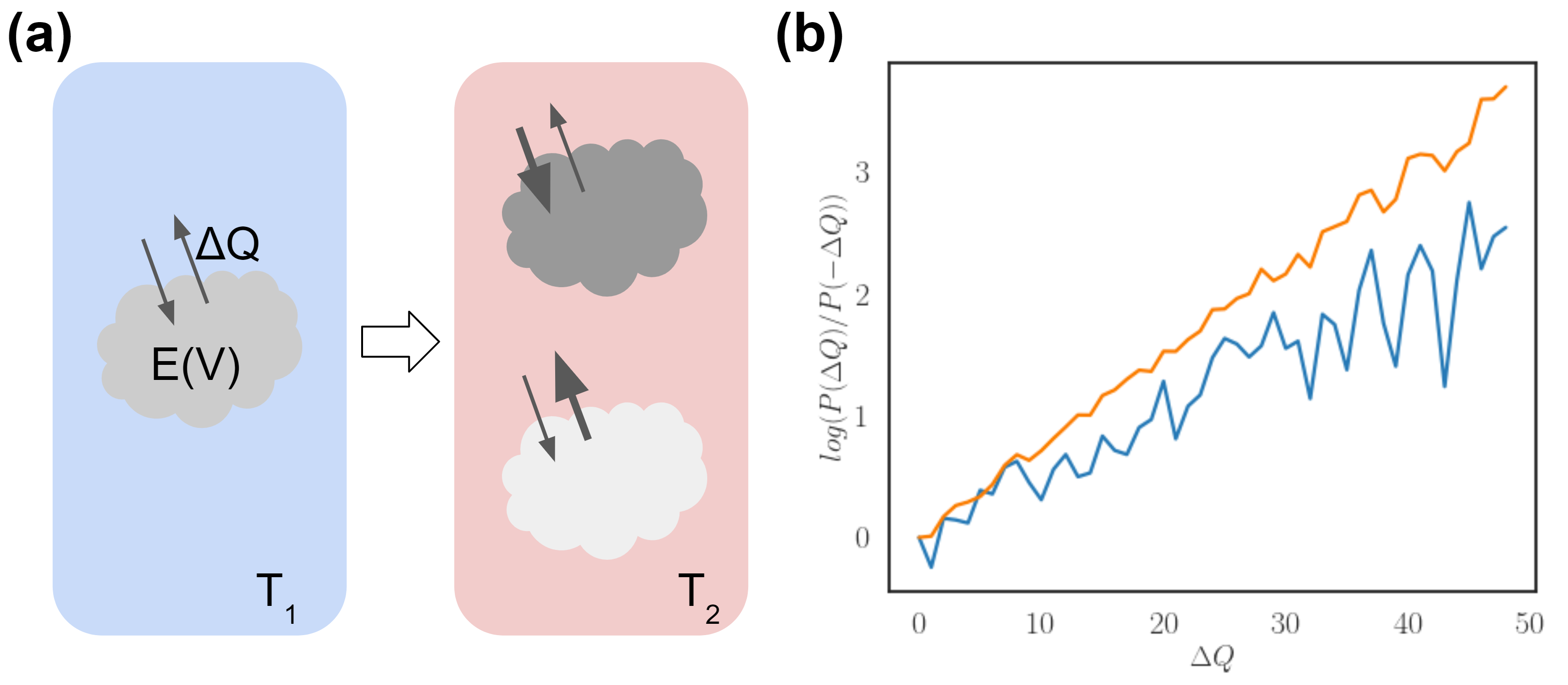}
    \caption{Testing XFT. (a) The experimental protocol of equilibrating the network (grey cluster) at $T_1$ (blue chart), then bringing it to $T_2$ (red chart) reservoir, with measurement of heat exchange $\Delta Q$ (grey arrows). The probability of $\Delta Q$ heat exchange values are recorded. (b) Heat exchange plotted as a function of the ratio of influx and out-flux heat probability. Temperature change for two curves are from $T_1=15$ to $T_2=5$ (orange) and from $T_1=15$ to $T_2=10$ (blue).}
    \label{fig5}
\end{figure}

Another variation of the fluctuation theorem focuses on the exchange of heat between two systems. We can extend our thermodynamics analysis for neural networks to a condition not driven by external input, but connected to another temperature after equilibrium, similar to a global modulation in the neural network (Fig. \ref{fig5}a). This corresponds to XFT that has a similar form as the original fluctuation theorem: $\frac{P(+\Sigma)}{P(-\Sigma)} = \Delta \beta \Sigma$, where $\Sigma$ is the measurement of entropy production. XFT replaces $\Sigma$ with heat flow $Q$ defined by equation \ref{heat} and can simply be replaced by energy $E$ when no work is performed. We show that the Hopfield network produces heat distribution following XFT (Fig. \ref{fig5}b).

\subsection{\label{sec:level1}Effects of biophysical constraints}
The network with Hebbian learning has connectivity matrix $W_{ij}$ that changes according to the ongoing input pattern $I_i(t)$. This process effectively accelerates the rate of convergence towards an equilibrium point in the energy landscape during our work protocol, since we design the input pattern to be a target fixed point. A changing energy landscape makes the definition of the true equilibrium free energy landscape ill-posed \cite{Yan2013-os}, but we argue that the nonequilibrium measurement holds as long as the learning dynamics is much slower than work protocols.

\begin{figure}[b!]
    \centering
    \includegraphics[width=0.5\textwidth]{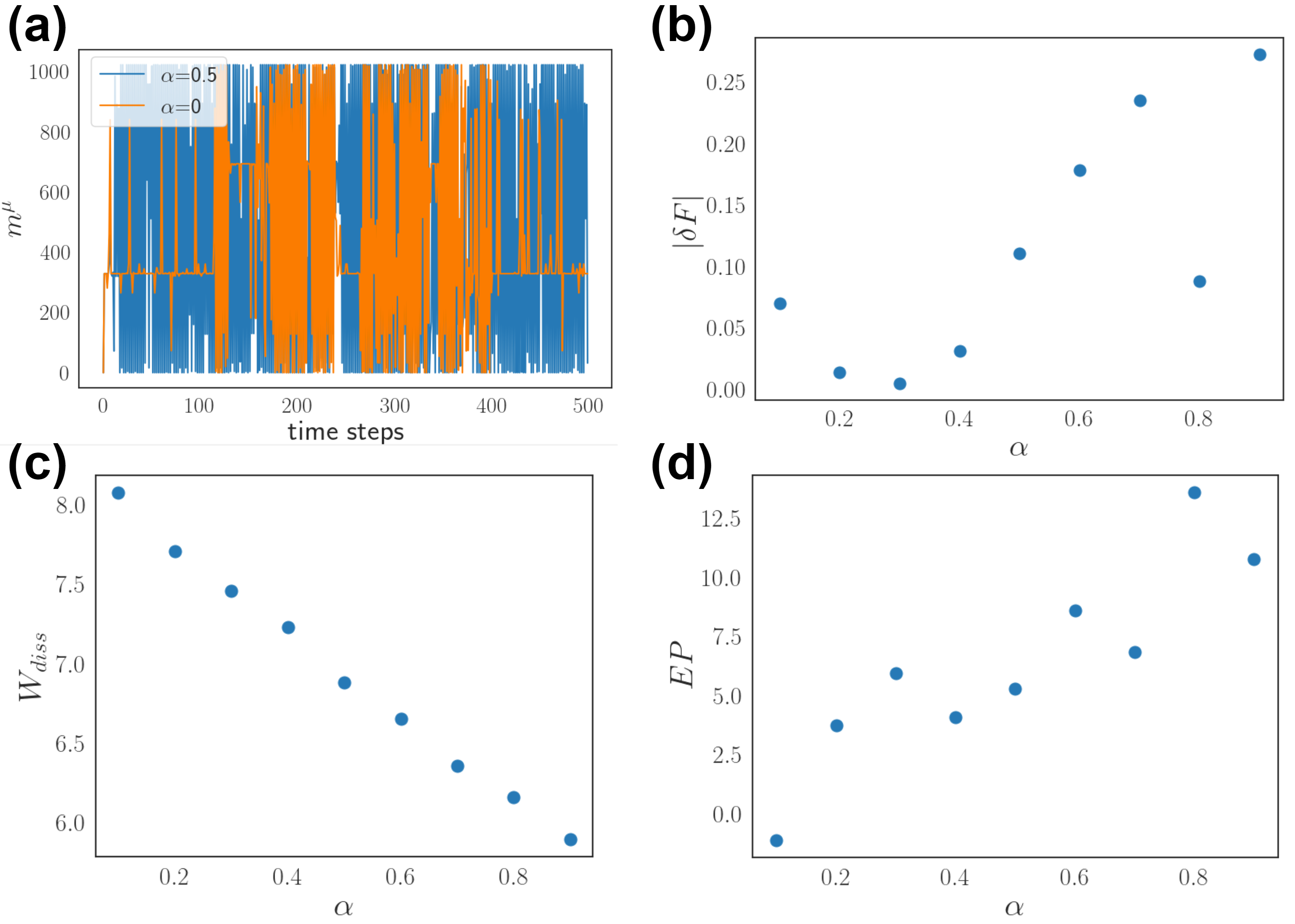}
    \caption{Effects of asymmetric connections. (a) Time series of neural states $m^{\mu}$ in neural networks with different asymmetric strengths $\alpha$. (b) The absolute error of free energy estimation $|\delta F|$ as a function of parameter $\alpha$. (c) The dissipated work $W_{diss}$ as a function of parameter $\alpha$. $\gamma =0.2$ in all simulations, resulting with true $\Delta F$ at the scale of $1$ for all instantiates. (d) Entropy production (EP) of networks as a function of asymmetric parameter $\alpha$.}
    \label{fig6}
\end{figure}

We explore two more constraints for biophysically plausible recurrent neural networks. One is the threshold of activation \cite{Hopfield1984-je} and the other is asymmetric connections \cite{Yan2013-os} (Fig. \ref{fig6}). The activation threshold $U$ has a strait forward definition in the energy function equation \ref{energy}. We tune this parameter and also show that CFT holds when it alters the free energy difference within a range. Note that in models with graded activities shown in equation \ref{continue}, the activation function may not be a step function but a monotonic and continuous function. The slope and turning point of this function acts similar to the effective temperature and threshold $U$ explored here (Fig. \ref{fig7}). 

Different from the Hopfield model, biological neural networks have directed synaptic connections and form asymmetric connectivity. The parameter $\alpha$ in equation \ref{ww} controls the weight of the cross over between two memory states. This can be thought as building an intermediate state between two attractors in the energy landscape. An important outcome of this term is that the connectivity matrix $W_{ij}$ may not be symmetric as in the original Hopfield model. The energy function would no longer be well-defined, since the derivative of energy functions would not guarantee to be pure gradient force (matrix $W_{ij}$ is not symmetric and may no longer positive semi-definite) \cite{Yan2013-os}. However, we are agnostic to this outcome and still proceed to analyze the nonequilibrium work distributions of the system (Fig. \ref{fig6}). The results hold for networks with asymmetric connections as long as the distribution of work protocols overlap and are sufficient for statistical sampling. When $\alpha$ is comparable to $\gamma$ or $1-\gamma$ that weights one of the attractor sates, the error of estimating free energy difference is small (Fig. \ref{fig6}b). As $\alpha$ increases, the dissipated work $W_{diss}$ for forward trajectories decreases and the entropy production from "housekeeping" heat $Q_{hk}$ increase (Fig. \ref{fig6}c,d). Here we quantify this part of entropy production with:

\begin{equation}\label{Qhk}
Q_{hk} = \log \big( \frac{P_{ss}(\epsilon_1) M(\epsilon_2|\epsilon_1)}{P_{ss}(\epsilon_2) M(\epsilon_1|\epsilon_2)} \big)
\end{equation}

where $P_{ss}$ is the steady-state probability of a given neural state and $M$ is the transition probability between states measured from steady-state \cite{Riechers2016-bv}. With the original symmetric network $\alpha=0$, detailed balance holds and this term would be zero. In other words, the entropy production here captures how irreversible the system is. This result is expected as the increase in asymmetric weight $\alpha$ on the crossover term results in a more irreversible process and the flux between state reduces dissipated work required. We notice that the notion of equilibrium free energy is inaccurate with calculation of equation \ref{energy} when connectivity $W_{ij}$ is asymmetric. However, results of recovering these free energy values still work in a reasonable range of asymmetry \cite{Crisanti1987-zj}. Further correction can be done by using steady-state values to compute non-equilibrium free energy \cite{Parrondo2015-zw,Yan2013-os}.

\section{\label{sec:level4}Discussion}
Biological networks are often driven by time-varying external input while maintaining internal structure such as memory patterns stored in neural networks \cite{Tank1987-oe, Ferrari2018-ll}. The resulting far-from-equilibrium trajectories of these driven systems can not be analyzed by methods based on equilibrium statistics \cite{Cessac2019-of, Yan2013-os, Zhong2018-vz}. In this manuscript, we introduce nonequilibrium methods of recovering equilibrium free energy difference in recurrent neural networks with a defined energy function. We show that the fluctuation theorem holds in our network model and further discuss how biophysical parameter alter the results.

Fluctuation theorem is one of the few descriptions in statistical mechanics that holds in far-from equilibrium conditions \cite{Crooks1999-zg}. The theorem relates distribution of nonequilibrium trajectories to equilibrium thermodynamic variables such as heat flux and free energy difference \cite{Jarzynski2011-lw}. Empirical evidence to corroborate fluctuation theorems started with single-molecule experiments, where work is intuitively defined by force generated by optical tweezers across certain length of a biological polymer \cite{Collin2005-fc}. The intersection of work required during forward and reversal pulling processes matches the equilibrium free energy difference of two molecular configuration. In addition, this phenomenon is observed in molecular dynamics simulations was well as single ion channel pulling experiments \cite{Cetiner2020-ci}. Together, these results show that nonequilibrium methods provide alternative ways of extracting equilibrium thermodynamic properties.

Similar ideas have been introduced to neural networks recently \cite{Yan2013-os, Zhong2018-vz, Roudi2011-hc, Monteforte2012-us}. In a stochastic thermodynamics setting, the change in weights of an perceptron during supervised learning can be defined as a forcing term and the objective function being an effective energy function \cite{Goldt2017-iy}. The thermodynamic efficiency during learning processes can then be defined in terms of the heat production due to weight updates and mutual information between the target signal and predictive output. In a deep unsupervised learning setup, convolution neural networks are trained to recover data destroyed through forward diffusion processes, showing how ideas of reversibility can applied to robust generative models \cite{Sohl-Dickstein2015-dm}. On the other hand, nonequilibrium thermodynamics can be used to analyze unsupervised learning process of a restricted Boltzmann machine (RBM) \cite{Salazar2017-xr}. The results show that fluctuation theorem holds in this abstract model and the learning process is related to the reduction of dissipated work. Interestingly, if we inspect the Hopfield's original simplified model with dimension analysis, one would conclude that the "energy" term has a unit that is closer to entropy production rate in nonequilibrium systems \cite{Chang1989-pm}. However, nonequilibrium thermodynamics investigation is still relatively absent in biological network models.

In order to make prediction for actual biological systems, it is less clear in these aforementioned work for how the methods can be applied to networks with recurrent structures or biophysically plausible constraints. Maximum entropy models have been one successful method used to describe statistics of neural population activity \cite{Schneidman2006-zt, Ferrari2018-ll}. The pair-wised interaction model, namely the Ising model, has a similar form as the Hopfield energy model. Related methods has been recently used to capture spatial temporal patterns by including activity history in a time window \cite{Mora2015-qb,Presse_Steve2013-ad, Pillow2008-vg}, but it is less clear how the parameters change under nonequilibrium settings. One recent study derives the nonequilibrium Green's function of neural networks and attempts to characterize functional connectivity with its time-dependent response to impulse \cite{Randi2020-mk}. The relation between this method and fluctuation theorem can be analyzed in future studies. It is known that glassy systems violate fluctuation dissipation theorem, thereby diminishing linear response theorem applied near equilibrium \cite{Marinari1998-xg}. More methods incorporating stochastic and nonequilibrium thermodynamics are required.

\begin{figure}[t!]
    \centering
    \includegraphics[width=0.5\textwidth]{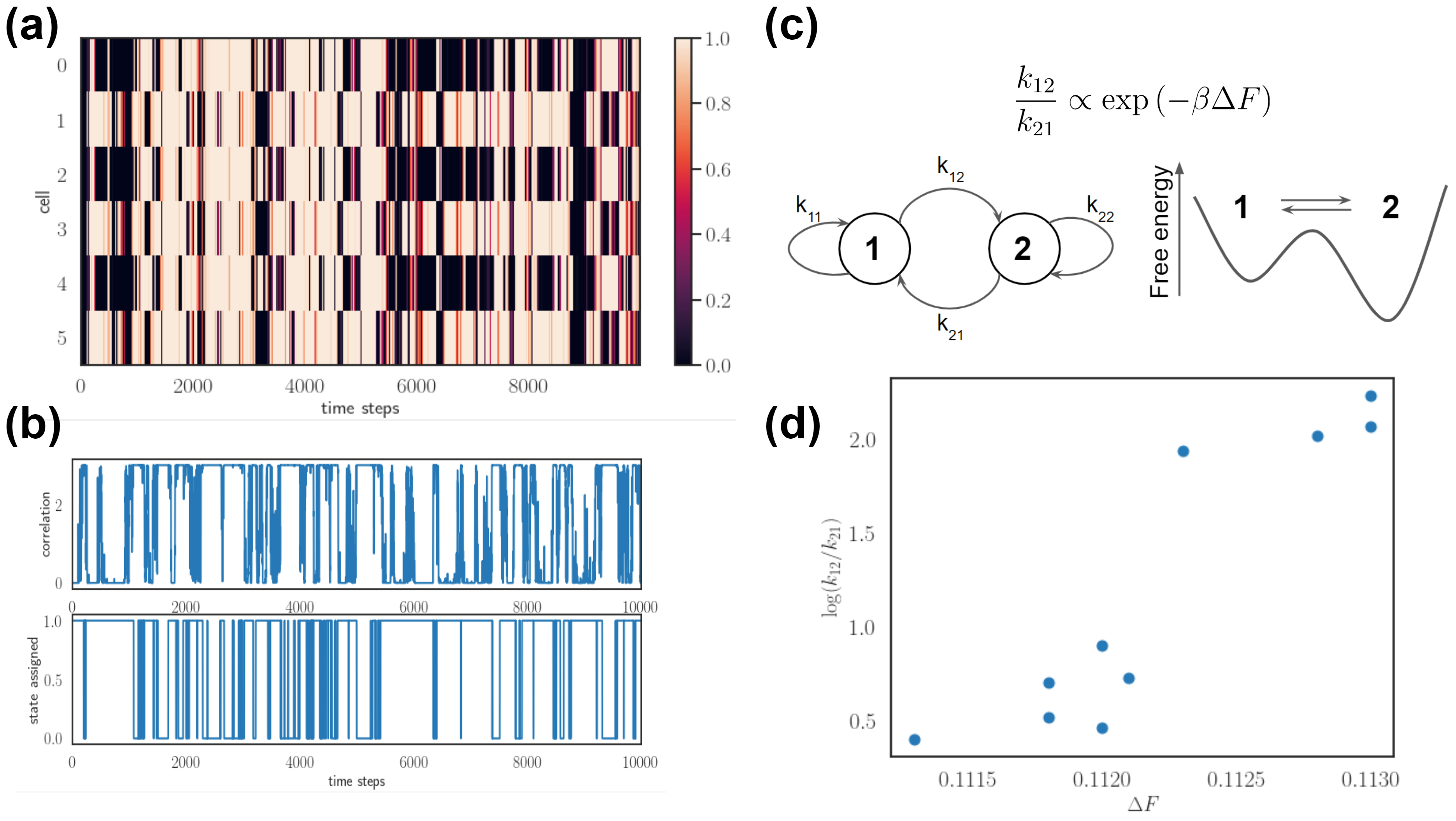}
    \caption{Free energy difference governs spontaneous neural dynamics. (a) Heat map of population firing rate through time ($N$=6). (b) Top: correlation with one of the $\mu$ patterns $\langle \epsilon^\mu r(t) \rangle$ through time. Bottom: Result of two-state transitions from HMM fit to the correlation time series (effectively assigning discrete states $m^\mu$ in the rate model). (c) Linking state transition to free energy difference $\Delta F$. The two-state transition kinetics (left) is reflects the free energy difference (right) between two states. (d) The true free energy difference plotted against the log ratio of forward and backward state transition kinetic coefficients.}
    \label{fig7}
\end{figure}

Biological networks have strong recurrent structures, sparse interactions, and heterogeneous physical properties \cite{Cessac2019-of, Vogels2005-sz}. We explore the recurrent property with the Hopfield network and tune the sparsity of interaction by adjusting memory patterns stored in the network or the activation threshold. We further relax the assumption of symmetric interaction by introducing the crossover term in the energy function. Another unexplored property is Dale's rule, which states that the sign of each neuron has to be conserved. This constraint leads to rich dynamics and balancing between excitatory and inhibitory activities in network dynamics \cite{Frady2019-in, Sompolinsky1988-yy}. We show that fluctuation theorem holds when some of these constraints are introduced. This opens a possibility of nonequilibrium measurements in realistic neural networks. For instance, an experimental prediction would be to compute "work" require in neural control experiments. The "work" in these systems are defined by the integral of input pattern along the ongoing activity patterns in the neural network. The recovered equilibrium free energy difference predicts the spontaneous transition rate between two target patterns across the forward and reversal protocol. More specifically, the equilibrium free energy difference links to reaction theory, according to Arrhenius equation: $K\propto \exp{(-\beta \Delta F)}$, where the equilibrium constant $K$ is the ratio of transition rate $k_{12}$ and $k_{21} $between two states (Fig. \ref{fig7}). Given the distribution of work, free energy measurements, and transition probabilities, one can also calculate dissipated work and entropy production in real neural activities as we demonstrated in asymmetrical networks. It is an empirical question to apply fluctuation theorem to nonequilibrium neural trajectories under external stimuli, recovering its equilibrium properties, then further predict its dynamical properties.

While we present simple forward and backward protocols for external stimuli, closed-loop experiments with complex stimulation in systems neuroscience has become a rising research direction \cite{Grosenick2015-fp}. In a closed-loops setup, the input pattern can depend on the network activity in real time. A notion of information measurement plays a role in the fluctuation theorem when the stimulus is not independent to neural activity. According to the thermodynamics of information, we can rewrite the Jarzynski equality as $\langle\exp{(-\beta W)} \rangle = \exp{(-\beta \Delta F - \Delta MI)}$, where $\Delta MI$ is the mutual information measured between neural activity and input protocol $MI(V,I)$ \cite{Parrondo2015-zw}. Mutual information between two random variables $X$ and $Y$ is calculated from the difference in entropy once an variable is observed: $MI(X,Y)=S(X)-S(X|Y)$, where $S$ is entropy defined in the former section. Since our stimulation protocol is independent to the neural activity in our simulations, $MI(V,I)$ is negligible (on the order of statistical error when we numerically compute the value). When this information term dominates, generalized fluctuation theorem should include the mutual information term and one can verify the theory through analyzing the distribution of forward and backward work distribution along with the information measurement. Note that this quantity of $MI(V,I)$ serves as a lower bound of dissipated work and a designed protocol with higher efficiency minimizes dissipated work by maximizing the predictive information during the work protocol \cite{Still2012-lf, Parrondo2015-zw}. In addition, mutual information between stimulus and neural response in the network has been used to quantify neural encoding in neuroscience \cite{Takagi2019-yz, Pillow2008-vg}. Whether or not the thermodynamic efficiency relates to the encoding performance is the goal for future directions.

\bibliography{NE_ref.bib}

\end{document}